\documentclass[12pt]{article}

\usepackage{mathrsfs}
\usepackage{amsmath}

\global\arraycolsep=1pt \oddsidemargin .20in \evensidemargin .5in
\usepackage[Symbol]{upgreek}
\usepackage{bbm}
\usepackage{dsfont}
\usepackage{amssymb}
\usepackage{textcomp}
\usepackage{caption}

\newcommand{\CR}{\nonumber \\*}

\newcommand{\trace}{\hbox {Tr}~}

\DeclareMathAlphabet{\mathpzc}{OT1}{pzc}{m}{it}

\def\N{\mathcal{N}}

\def\ga{\gamma}

\def\susy {\delta^{\scriptscriptstyle \,Susy}}
\def\gauge{\delta^{\scriptscriptstyle \,gauge}}

\def\be{\begin{equation}}
\def\ee{\end{equation}}
\def\bea{\begin{eqnarray}}
\def\eea{\end{eqnarray}}
\def\bdis{\begin{displaymath}}
\def\edis{\end{displaymath}}
\def\nn{\nonumber}

\def\A{\mathbb{A}}

\newcommand{\scal}[1]{\bigl ({#1} \bigr )}

\def\A{\mathscr{A}}
\def\B{\mathscr{B}}

\def\susy {\delta^{\scriptscriptstyle \,Susy}}
\def\gauge{\delta^{\scriptscriptstyle \,gauge}}

\def\s{\,\mathpzc{s}\,}

\def\zero{{\scriptscriptstyle \rm c}}
\def\un{{\scriptscriptstyle \rm inv}}

\def\a{{\scriptscriptstyle (\mathpzc{s})}}
\def\q{{\scriptscriptstyle (Q)}}

\def\S{{\mathcal{S}_\a}}
\def\Q{{\mathcal{S}_{\q}}}

\begin{document}
\allowdisplaybreaks[1]
\renewcommand{\thefootnote}{\fnsymbol{footnote}}
\def\corr{$\spadesuit $}
\def\trefle{$\clubsuit$}

\renewcommand{\thefootnote}{\arabic{footnote}}
\setcounter{footnote}{0}

\begin{titlepage}
\renewcommand{\thefootnote}{\fnsymbol{footnote}}
\begin{flushright}
\
\vskip -3cm
{\small CERN-PH-TH/2008-192}\\
\vskip 3cm
\end{flushright}
\begin{center}
{{\Large \bf

Supersymmetric Adler--Bardeen anomaly  in~$\N=1$~super-Yang--Mills theories

 }}
\lineskip .75em
\vskip 3em
\normalsize
{\large Laurent Baulieu\footnote{email address: baulieu@lpthe.jussieu.fr},
Alexis Martin\footnote{email address: alexis.martin@lpthe.jussieu.fr}\\
\vskip 1em
$^{* }${\it Theoretical Division CERN} \footnote{ CH-1211 Gen\`eve, 23, Switzerland}
\\
$^{* \dagger}${\it LPTHE, CNRS and Universit\'e Pierre et Marie
Curie } \footnote{ 4 place Jussieu, F-75252 Paris Cedex 05,
France}
}

\vskip 1 em
\end{center}
\vskip 1 em
\begin{abstract}
We provide a   study of the supersymmetric Adler--Bardeen anomaly in the  $\N=1,\ d=4,6,10$ super-Yang--Mills theories.
We work in the component formalism that includes shadow fields, for  which Slavnov--Taylor identities can be independently set for both gauge invariance and supersymmetry. We find  a method with improved descent equations  for getting the solutions of   the  consistency conditions of both Slavnov--Taylor identities  and finding  the local field  polynomials for the standard Adler--Bardeen anomaly and    its supersymmetric counterpart.  We
  give   the  explicit solution    for the ten-dimensional case.
\end{abstract}

\end{titlepage}
\renewcommand{\thefootnote}{\arabic{footnote}}
\setcounter{footnote}{0}

\section{Introduction}
The  gaugino   of even-dimensional  $\N=1$  supersymmetric Yang--Mills theories  is
a chiral    spinor. This implies the existence of an  Adler--Bardeen     one-loop anomaly. Its effect   is made manifest by
the  non-vanishing of a relevant form factor of well defined  one-loop amplitudes,   as predicted by
the consistency equations  \cite{Wess.1971} and  by  their solution given by the Chern--Simons formula   \cite{Stora.1984,Baulieu.1984,Zumino.1984}.
It is of course well-known that the existence of anomalies in ten-dimensional supersymmetric Yang--Mills
theory has triggered fundamental progress in string theories    \cite{Witten.AlvarezGaume,Green.1984}.
The consistency of these anomalies with~$\N=1$  supersymmetry eluded however a complete analysis.   In
this paper, we address this problem using recent progress in the formulation of supersymmetric theories
in component formalism and in various dimensions.

  As a matter of fact, in  a supersymmetric gauge theory,
  the Wess and Zumino consistency conditions must be generalized  in order
  to be compatible with supersymmetry. The  standard Adler--Bardeen anomaly must come
  with a supersymmetric counterpart. The method of~\cite{Stora.1984,Baulieu.1984,Zumino.1984}
  was however generalized to higher dimensions both in component formalism \cite{Nair.1985} and in superspace~\cite{Bonora.1986} to determine
  this supersymmetric counterpart, but no explicit expression was given for the
  ten-dimensional case. Such an expression was derived later in~\cite{Candiello.1994}
  for the coupled $\N=1$ supergravity and super-Yang--Mills theory for the supersymmetrization
  of the Green--Schwartz mechanism~\cite{Green.1984}.

On the other hand, recent results in component formalism based on the introduction of
shadow fields \cite{shadow} have allowed for the definition of two independent Slavnov--Taylor identities.
This    has  permitted  the  disentangling of   gauge invariance and supersymmetry.
In this way, one gets a consistent analysis of the   compatibility of the Adler--Bardeen
anomaly with  supersymmetry, and an algebraic proof was given for the absence of anomalies
in~$\N=2,4,\ d=4$ super-Yang--Mills theories and  for the fact  that in the case of $\N=1$
the only possible anomaly is of the Adler--Bardeen type.

The purpose of this paper is to give a systematic way for solving the
supersymmetric consistency equations for the supersymmetric Adler--Bardeen
anomaly in the cases of   $\N=1,\ d=4,6,10$ super-Yang--Mills theories, following
the same logic as that of \cite{shadow} and completing it by determining the explicit
expression for the ten-dimensional case.

\section{Adler--Bardeen anomaly in super-Yang--Mills theories}
We first introduce some definitions that apply to $\N=1$ supersymmetric Yang-Mills theories in general and focus on the ten-dimensional case afterwards.
Let $\s$ be the BRST operator associated to ordinary gauge
symmetry and $Q$ the  differential operator that acts on the physical fields
as an ordinary supersymmetry transformation minus a gauge transformation of
parameter a scalar field $c$, that is $Q\equiv\susy-\gauge(c)$. The shadow field $c$  allows for the elimination of the field dependent gauge transformations in the commutators of the supersymmetry algebra \cite{shadow}. It completes  the usual Faddeev--Popov ghost $\Omega$ associated to BRST symmetry. The $\s$ and $Q$ operators verify
\be \label{tzatziki}
\s^2=0\, , \quad \{\s,Q\}=0\, , \quad Q^2\approx \mathcal{L}_\kappa
\ee
where $\approx$ means that this relation can hold modulo the equations of motion and $\kappa$ is the  bilinear function of the supersymmetry parameter, $\kappa^\mu=-i(\epsilon\gamma^\mu\epsilon)$. In addition to the ghost number, we assign a shadow number,  equal to one for the supersymmetry parameter and for the shadow field $c$, and zero for the other fields. The $Q$ operator increases the shadow number by one unit. Each field and operator  has a grading determined by the sum of the ghost number, shadow number and   form degree. Transformation laws for the various fields can be deduced from the definition of an extended curvature $\tilde{F}$, by decomposition over terms of all possible   gradings of  the following horizontality condition
\be \label{courbure N1 anomalie generique}
\tilde{F} \equiv (d + \s + Q - i_\kappa)
\scal{ A + \Omega + c} + \scal{ A + \Omega
 + c}^2 = F + \susy A
\ee
where $A$ is the gauge connection and $F=dA+AA$.
At the quantum level,  one
introduces    sources for the   non-linear  $\s$, $Q$ and $\s Q$ transformations of all fields.
The   BRST invariant gauge-fixed local action with all needed external sources is then  given by
\be \label{action classique}
\Sigma = S[\varphi] +\s\Uppsi + S_{\rm ext}
\ee
The   BRST  and supersymmetry      invariances of $\Sigma$ imply both    Slavnov--Taylor identities
\be \label{Slav class}
\S(\Sigma)=0\, , \quad \Q(\Sigma)=0
\ee
where $\S$ and $\Q$ are the Slavnov--Taylor operators associated to the $\s$ and $Q$ operators, respectively \footnote{We refer to \cite{shadow} for   more  explicit definitions.}.
These identities  imply the following anticommutation relations between the associated linearized Slavnov--Taylor operators $\S_{|\Sigma}$  and $\Q_{|\Sigma}$
\be \label{nilpot LST}
\S_{|\Sigma}^2=0\, , \quad \{\S_{|\Sigma}\,, \Q_{|\Sigma}\}=0\, , \quad \Q_{|\Sigma}^2=\mathcal{P}_\kappa
\ee
where $\mathcal{P}_\kappa$ is the differential operator that acts as the Lie derivative along $\kappa$ on the fields and external sources \footnote{The fact that $Q^2$ is a pure derivative only modulo the equations of motion on the gaugino of the ten-dimensional case is solved for the linearized Slavnov--Taylor operator $\Q_{|\Sigma}$ by introducing suitable source terms in (\ref{action classique}).}.
 An anomaly is   defined as an obstruction -- at a certain order $n$ of perturbation -- to the implementation of the Slavnov--Taylor identities on the vertex functional $\Gamma=\Sigma+O(\hbar)$, that is
\be \label{Slav break}
\S(\Gamma)=\hbar^n\A\, , \quad \Q(\Gamma)=\hbar^n\B
\ee
where $\A,$   and  $    \B$ are   respectively        integrated local functionals of ghost number one  and shadow number one,     defined modulo    $\S$- and  $\Q$-exact terms. The introduction of the       linearized Slavnov--Taylor operators permits one to   write  the consistency conditions
\be
\S_{|\Sigma}\A=0^{(2,0)}\, , \quad \Q_{|\Sigma}\A +\S_{|\Sigma}\B=0^{(1,1)}\, , \quad    \Q_{|\Sigma}\B=0^{(0,2)}
\ee
where the superscripts $(g,s)$ denote the ghost and the shadow number.
Due to (\ref{nilpot LST}), the problem of the determination of the solutions to these conditions is   a cohomological problem. The consistent Adler--Bardeen anomaly is thus    defined as the pair $\A$ and  $\B$, identified as   the  elements   $(1,0   )$ and     $(0,1  )$    of the cohomology of the   operators     $\S_{|\Sigma}, \Q_{|\Sigma}$, in the set of integrated local functionals depending on the fields and sources.
It can be shown that the cohomology of the linearized Slavnov--Taylor operators in the set of local functionals depending on the fields and sources is completely determined by that of the classical operators in the set of local functionals depending only on the fields, provided such functionals are identified on the stationary surface, i.e., modulo equations  of motion~\cite{Henneaux.1992,shadow}.
We will thus consider the consistency conditions
\be \label{ce} \s \A =
0^{(2,0)},\quad Q\A+\s \B = 0^{(1,1)},\quad Q \B =0^{(0,2)}
\ee

To  determine the solutions  of these  equations    in $d=2n-2$ dimensional space-time, we formally define the Chern character $2n$-form  ${\rm Ch}_{n}\equiv\trace \tilde{F}^n$,
where      $\tilde F$  has been introduced  in Eq.~(\ref{courbure N1 anomalie generique})       \footnote{The following procedure is actually valid for any invariant symmetric polynomial, which covers the case of so-called factorized anomalies.}.   From a generalization of the algebraic Poincar\'e lemma  and   the Chern--Simons identity,      ${\rm Ch}_{n}$ can locally be written as a $(\tilde{d}\equiv d+\s+Q-i_\kappa)$-exact term
\be \label{CS 12}
\trace \tilde{F}^n = \tilde{d}\,\trace W_{2n-1}(\tilde{A},\tilde{F})
\ee
$W_{2n-1}$ is the Chern--Simons form, which can be calculated form the formula
\be
W_{2n-1}(\tilde{A},\tilde{F}) = n\!\int_0^1 {\rm d}t\,\trace(\tilde{A}\,\tilde{F}_t^{\,n-1})
\ee
where $F_t=tdA+t^2A^2$.
The term with grading $(2,0)$ in (\ref{CS 12}) gives the standard Adler--Bardeen anomaly \cite{Zumino.1984}
\be
\A \equiv \int W_{2n-2}^{(1,0)}\, , \quad \s\A=0
\ee
The term with grading $(1,1)$ gives a solution for the consistency condition
\be \label{QA sB}
Q\A +\s\mathscr{B}^\zero=0
\ee
which is given by
\be
\mathscr{B}^\zero \equiv\int W_{2n-2}^{(0,1)}
\ee
However, we have not yet a solution to the consistency equations, since the term  $\trace   \tilde{F}^n $     with grading $(0,2)$
  gives a  breaking of the   consistency condition  $Q \mathscr{B}^\zero  =0   $, according to
\be\label{truc}
Q \mathscr{B}^\zero = {\scriptstyle\binom{n}{2}}\int \trace \susy A\,\,\susy A\,\, F^{n-2}
\ee
where ${\scriptstyle\binom{n}{2}}$ stands for the binomial coefficient.

The solution of this problem can be solved as follows.  One observes that   $\mathscr{B}^\zero$
in Eq.~(\ref{truc})
is a particular solution of Eq.~(\ref{QA sB}),    so that  one can add to it a local functional of the fields $\mathscr{B}^\un$, provided their sum is $Q$-invariant. To preserve the condition~(\ref{QA sB}), $\mathscr{B}^\un$ must be $\s$-closed. But since  $Q\mathscr{B}^\zero$ is not $\s$-exact and  since  $\{\s,Q\}=0$, no $\s$-exact element of $\mathscr{B}^\un$  can contribute and   $\mathscr{B}^\un$ must be in the cohomology of $\s$. Therefore, the consistency conditions (\ref{ce}) are fulfilled provided there exists a gauge-invariant local functional of the physical fields satisfying
\be \label{kalymnos}
\susy \mathscr{B}^\un = - {\scriptstyle\binom{n}{2}}\int \trace \susy A\,\,\susy A\,\, F^{n-2}
\ee
so that
\be
\mathscr{B} = \mathscr{B}^\zero + \mathscr{B}^\un\, , \quad Q\mathscr{B}=0
\ee

We now address the problem of determining   $\mathscr{B}^\un$.
We  keep general $d$-dimensional notations, since we  have in mind the cases of   $\N=1$ super-Yang--Mills theories   in $d=4,6$ and~$10$. The field content is made of a gauge connexion $A=A_\mu dx^\mu$ ($\mu=0,\ldots,d$) and its gaugino $\lambda$, both in the adjoint representation of some gauge group. Transformation laws are determined by Eq.~(\ref{courbure N1 anomalie generique}) and its Bianchi identity with $\susy A=-i(\epsilon\gamma_1\lambda)$, and $\gamma_1\equiv\gamma_\mu dx^\mu$. We take $\epsilon$ commuting so that  Eq.~(\ref{tzatziki}) holds. To determine $\mathscr{B}^\un$,    we first make  the following observation. In each of the considered dimensions, a Fierz identity shows that $\kappa^\mu\equiv-i(\epsilon\gamma^\mu\epsilon)$ is light-like, that is $\kappa^\mu\kappa_\mu=0$. We then introduce a vector $\hat\kappa^\mu$ that we normalize so that $\hat\kappa^\mu\kappa_\mu=1$. Let moreover $\iota_\kappa$ be the contraction operator along $\kappa^\mu$, so that
\be
 \susy A = -i(\epsilon\gamma_1 \lambda) \, ,\quad \susy F= -d_A\susy A \, ,\quad \iota_\kappa\susy A =0\, ,\quad (\susy)^2 A = \iota_\kappa F
\ee
By integrating by parts and with the Bianchi identity $d_A F=0$, it is straightforward to see that the following expression
\be \label{Geraka}
\mathscr{B}^\un =c_n \int\trace\Bigl(\hat\kappa\,\susy A \,\susy A \,\susy A \,F^{\,n-3} \Bigr)
\ee
with $\hat\kappa\equiv\hat\kappa_\mu dx^\mu$ and $c_n=\frac{n-2}{3}{\scriptstyle\binom{n}{2}}$ is such that     Eq.~(\ref{kalymnos}) holds true. Moreover, it provides an off-shell expression, as it is solely based on the geometrical curvature Eq.~(\ref{courbure N1 anomalie generique}) and its Bianchi identity. The problem thus reduces to that of the elimination of $\hat\kappa$, in order the solution to be bilinear in the supersymmetry parameter. In four dimensions for example, $n=3$ and the elimination of $\hat\kappa$ in Eq.~(\ref{Geraka}) directly yields the known result \cite{shadow}.

   From now on, we focus on the ten-dimensional super-Yang--Mills theory. Its fields content consists of a gauge connection $A=A_\mu dx^\mu$ ($\mu=0,\ldots,9$) and a Majorana--Weyl spinor $\lambda$, with both $\lambda$ and $\epsilon$ $\in {\bf 16_+}$ of $SO(1,9)$. Eq.~(\ref{kalymnos}) now reads
\be \label{kalymnos2}
\susy \mathscr{B}^\un = - 15\,\int \trace \susy A\,\,\susy A\,\, F^4
\ee
By demanding removal of the $\hat\kappa$ dependency in Eq.(\ref{Geraka}), one is naturally led to consider the following solution
\be
\mathscr{B}^\un = \frac{1}{16}\int {\rm d}^{10}x\,\trace\Bigl( \varepsilon^{\mu_1\cdots\mu_{10}}(\epsilon\gamma_{\mu_1\mu_2}^{\phantom{\mu_1\mu_2}\sigma}\lambda)(\lambda\gamma_{\mu_3\mu_4\sigma}\lambda)\,F_{\mu_5\mu_6}F_{\mu_7\mu_8}F_{\mu_9\mu_{10}}\Bigr)
\ee
Indeed, with the help of some ten-dimensional $\ga$-matrix identities [\ref{appendix}], one can check that modulo the equations of motion
\bea \label{calcul}
 \susy \int {\rm d}^{10}x\, \trace&& \Bigl( \varepsilon^{\mu_1\cdots\mu_{10}}(\epsilon\gamma_{\mu_1\mu_2}^{\phantom{\mu_1\mu_2}\sigma}\lambda)(\lambda\gamma_{\mu_3\mu_4\sigma}\lambda)\,F_{\mu_5\mu_6}F_{\mu_7\mu_8}F_{\mu_9\mu_{10}}\Bigr)\CR
&&\approx 15\int {\rm d}^{10}x\,\trace \Bigl( \frac{1}{16}\varepsilon^{\mu_1\cdots\mu_{10}}\,(\epsilon\gamma^\sigma\epsilon)(\lambda\gamma_{\mu_1\mu_2\sigma}\lambda)F_{\mu_3\mu_4}
F_{\mu_5\mu_6}F_{\mu_7\mu_8}F_{\mu_9\mu_{10}}\CR
&&\hspace{19mm}+\frac{1}{96}\varepsilon^{\mu_1\cdots\mu_{10}}\,(\epsilon\gamma_{\mu_1\mu_2}^{\phantom{\mu_1\mu_2}\nu_1\nu_2\nu_3}\epsilon)(\lambda\gamma_{\nu_1\nu_2\nu_3}\lambda)F_{\mu_3\mu_4}
F_{\mu_5\mu_6}F_{\mu_7\mu_8}F_{\mu_9\mu_{10}}\Bigr)\nn
\eea
\be
=-15\int {\rm d}^{10}x\,\trace \Bigl(\varepsilon^{\mu_1\cdots\mu_{10}}\,(\epsilon\gamma_{\mu_1}\lambda)(\epsilon\gamma_{\mu_2}\lambda)F_{\mu_3\mu_4}
F_{\mu_5\mu_6}F_{\mu_7\mu_8}F_{\mu_9\mu_{10}}\Bigr)
\ee
It implies that the  following expression\be \label{Falkonera}
\mathscr{B}=\int W_{10}^{(0,1)}+\frac{1}{16}\int {\rm d}^{10}x\,\trace\Bigl( \varepsilon^{\mu_1\cdots\mu_{10}}(\epsilon\gamma_{\mu_1\mu_2}^{\phantom{\mu_1\mu_2}\sigma}\lambda)(\lambda\gamma_{\mu_3\mu_4\sigma}\lambda)\,F_{\mu_5\mu_6}F_{\mu_7\mu_8}F_{\mu_9\mu_{10}}\Bigr)
\ee
solves the supersymmetric part of the consistency equations. We have therefore found    the  supersymmetric counterpart of the Adler--Bardeen anomaly for the  $\N=1,d=10$ super-Yang--Mills theory. The result can be easily transposed in $d=4$ and $6$ dimensions. For the case $d=4$, we recover the result of \cite{shadow}, where the problem is  less involved and can easily be  solved by   inspection over all possible field polynomials.

\subsection*{Acknowledgments}
We thank very much G.~Bossard for useful discussions. A.~M. is
grateful to P.~Vanhove for his useful advice regarding gamma
matrix manipulations. This work has been partially supported by
the contract ANR (CNRS-USAR), \texttt{05-BLAN-0079-01}. A.~M. has
been supported by the Swiss National Science Foundation, grant
\texttt{PBSK2-119127}.

\appendix

\section{Ten-dimensional $\ga$-matrix identities}\label{appendix}
The ten-dimensional $\ga$-matrices satisfy the Clifford algebra $\{\gamma^\mu,\gamma^\nu\}=2\eta^{\mu\nu}$ and our convention for antisymmetrization is $\gamma^{\mu_1\cdots\mu_n}=\frac{1}{n!}\gamma^{[\mu_1}\cdots\gamma^{\mu_n]}$. Both $\epsilon$ and $\lambda$ are chiral, so that we only have to consider a basis of gamma matrices made of
\be \label{base}
\gamma^\mu\, ,\quad \gamma^{\mu_1\mu_2\mu_3}\, , \quad \gamma^{\mu_1\cdots\mu_5}
\ee
Useful identities used to derive (\ref{calcul}) are
\be\begin{split}
&\ga^\sigma\ga^\mu\ga_\sigma=-8\ga^\mu\CR
&\ga^\sigma\ga^{\mu_1\mu_2\mu_3}\ga_\sigma=-4\ga^{\mu_1\mu_2\mu_3}\CR
&\ga^\sigma\ga^{\mu_1\cdots\mu_5}\ga_\sigma=0\CR\end{split}\hspace{10mm}\begin{split}
&\ga^{\sigma_1\sigma_2\sigma_3}\ga^\mu\ga_{\sigma_1\sigma_2\sigma_3}=288\ga^\mu\CR
&\ga^{\sigma_1\sigma_2\sigma_3}\ga^{\mu_1\mu_2\mu_3}\ga_{\sigma_1\sigma_2\sigma_3}=-48\ga^{\mu_1\mu_2\mu_3}\CR
&\ga^{\sigma_1\sigma_2\sigma_3}\ga^{\mu_1\cdots\mu_5}\ga_{\sigma_1\sigma_2\sigma_3}=0\CR\end{split}
\ee
as well as $\ga^{\sigma_1\cdots\sigma_5}\ga^\mu\ga_{\sigma_1\cdots\sigma_5}=\ga^{\sigma_1\cdots\sigma_5}\ga^{\mu_1\mu_2\mu_3}\ga_{\sigma_1\cdots\sigma_5}=\ga^{\sigma_1\cdots\sigma_5}\ga^{\mu_1\cdots\mu_5}\ga_{\sigma_1\cdots\sigma_5}=0$.
A generic bi-spinor can be expanded over the basis (\ref{base}) as
\be
\xi\zeta=\frac{1}{16}(\xi\ga^\sigma\zeta)\ga_\sigma +\frac{1}{96}(\xi\ga^{\sigma_1\sigma_2\sigma_3}\zeta)\ga_{\sigma_1\sigma_2\sigma_3}+\frac{1}{3840}(\xi\ga^{\sigma_1\cdots\sigma_5}\zeta)\ga_{\sigma_1\cdots\sigma_5}
\ee
In particular, the supersymmetry parameter $\epsilon$ being commuting, the non-vanishing terms for $\zeta=\xi=\epsilon$ are $(\epsilon\ga^\sigma\epsilon)$ and $(\epsilon\ga^{\sigma_1\cdots\sigma_5}\epsilon)$, so that for example $(\epsilon\ga^\sigma\epsilon)\epsilon\ga_\sigma=0$. $\lambda$ being anticommuting, the only non-vanishing term for $\zeta=\xi=\lambda$ is $(\lambda\ga^{\sigma_1\sigma_2\sigma_3}\lambda)$. The following identities also turned out to be precious
\bea
&&(\epsilon\ga_{\mu_1}\lambda)(\epsilon\ga_{\mu_2}\lambda)=-\frac{1}{16}(\epsilon\ga^\sigma\epsilon)(\lambda\ga_{\mu_1\mu_2\sigma}\lambda)-\frac{1}{96}(\epsilon\gamma_{\mu_1\mu_2}^{\phantom{\mu_1\mu_2}\nu_1\nu_2\nu_3}\epsilon)(\lambda\gamma_{\nu_1\nu_2\nu_3}\lambda)\CR
&&(\epsilon\ga^\sigma\lambda)(\epsilon\gamma_{\mu_1\mu_2\sigma}\lambda)=-\frac{3}{8}(\epsilon\ga^\sigma\epsilon)(\lambda\ga_{\mu_1\mu_2\sigma}\lambda)+\frac{1}{48}(\epsilon\gamma_{\mu_1\mu_2}^{\phantom{\mu_1\mu_2}\nu_1\nu_2\nu_3}\epsilon)(\lambda\gamma_{\nu_1\nu_2\nu_3}\lambda)\CR
&&(\epsilon\gamma_{\mu_1\mu_2}^{\phantom{\mu_1\mu_2}\nu_1\nu_2\nu_3}\lambda)(\epsilon\gamma_{\nu_1\nu_2\nu_3}\lambda)=-\frac{21}{4}(\epsilon\ga^\sigma\epsilon)(\lambda\ga_{\mu_1\mu_2\sigma}\lambda)-\frac{3}{8}(\epsilon\gamma_{\mu_1\mu_2}^{\phantom{\mu_1\mu_2}\nu_1\nu_2\nu_3}\epsilon)(\lambda\gamma_{\nu_1\nu_2\nu_3}\lambda)
\eea
Needless to say, Ulf Gran's GAMMA package \cite{Gran} was greatly appreciated to derive these identities.

\end{document}